# A Small-Eddy-Dissipation Mechanism for Developing Turbulence Models

Yan Jin

Institute of Multiphase Flows, Hamburg University of Technology, 21073, Hamburg, Germany, e-mail: yan.jin@tuhh.de

**Abstract**

Jin (*Phys. Fluids*, vol. 31, 2019, p. 125102) proposed a new turbulence simulation method which shows better performance than other classic turbulence models. It is composed of a small-eddy-dissipation mixing length (SED-ML) model for calculating the reference solution and a parameter extension method for correcting the solution. The mechanism of this method is more deeply analyzed in this study to find out how to develop a turbulence model with a high accuracy and a low computational cost. The turbulent channel flows with $Re_\tau = 821$ and 2003 and decaying homogenous and isotropic turbulence are simulated to demonstrate how the new turbulence simulation method works. The small-eddy-dissipation (SED) mechanism for developing turbulence models has been proposed through our analysis. According to this mechanism, the model solution is an asymptotic approximation of the exact solution of the Navier-Stokes equations. The modeling term introduces an artificial dissipation which dissipates small eddies. The purpose of turbulence modeling is to dissipate more small eddies without changing the statistical solution qualitatively. We expect more small eddies can be dissipated where the turbulence is stronger. This mechanism is different from RANS which approximates the Reynolds stresses and LES which approximates the sub-grid-scale (SGS) motions, while it interprets the physics of turbulence modeling more precisely. A modified mixing length with two damping functions is developed to identify the characteristic length of turbulence, leading to the SED-ML model. The simulation accuracy can be further improved using a linear extension. Our numerical results show that the SED-ML model is in accordance with the SED mechanism. This might explain why the new method is more accurate than RANS while it requires a lower computational cost than LES.

## 1. Introduction

Turbulence or turbulent flow is fluid motion characterized by chaotic changes in pressure and flow velocity (Pope, 2000). It is often observed in everyday phenomena and most fluid flows occurring in nature or created in engineering applications are turbulent.

Turbulent motions are governed by the Navier-Stokes equations. These equations can be solved directly; the corresponding simulation is called direct numerical simulation (DNS). DNS is the most accurate method for simulating turbulent flows, however, it is hard to use DNS to simulate turbulent flows with high Reynolds numbers because of its high computational cost. Turbulent flows in real industrial problems are more often simulated with the help of turbulence modeling.

The most popular mathematical methods for simulating turbulent flows are the Reynolds Averaged Navier-Stokes equations simulation (RANS) and large eddy simulation (LES). In RANS, the Reynolds stresses are modeled, often with the help of the eddy-viscosity assumption (EVA) which states that the anisotropic Reynolds stress tensor is proportional to the mean strain-rate tensor. Representative RANS models are the $k-\varepsilon$ and $k-\omega$ models, see Yakhot & Orszag (1986) and Menter (1994) as examples. Nevertheless, the EVA for RANS is not true even for simple turbulent flows, e.g., turbulent flows in a smooth wall channel (Pope, 2000). The Reynolds Stress Models (RSM) avoid using an isotropic eddy viscosity and describe the transport of each Reynolds stress, see Gibson & Launder (1978) as an example. However, Jin & Herwig (2015) showed that the RSMs are not more accurate than $k-\varepsilon$ and $k-\omega$ models when they are used to simulate the turbulent flows in a rough wall channel. The model errors of all these RANS models are higher than 20%.

In LES, the large eddies are directly resolved while small eddies are modeled. Most LES models try to approximate the fluid motions within the sub-grid scales (SGS), which are all scales that are smaller than the mesh size $\Delta$ (Smagorinsky 1963; Nicoud & Ducros 1999; Kim et al. 1997). This $\Delta$ in LES should lie in the inertial range of the energy spectrum. However, there is not an inertial range for weak turbulence at low or medium Reynolds numbers. Even if the flow has a high Reynolds number, one still cannot ensure that the SGS motions are precisely modeled because the SGS motions are related to the computational mesh, which is usually generated empirically. So, the interpretations of RANS and LES don't precisely reflect the physics of how they work. Due to the deficiencies in physical interpretation, it is hard to further improve the generality and accuracy and reduce the computational cost of the turbulence models.

Jin (2019) interpreted the turbulence modeling in a different way and proposed a new method, the parameter extension simulation (PES) method, for simulating turbulent flows.

Instead of modeling the Reynolds stressor term or the SGS term, the new method treats the turbulence modeling as an asymptotic approximation of the exact solution of the turbulent flows. A modified mixing length (ML+) model was developed to calculate the reference solution. The accuracy of the statistical results can be further improved using a linear extension from the reference solution.

The PES has been used to simulate various turbulent flows for validation, including the rough wall channel flows and compressor cascade flows with high Reynolds numbers. The numerical results show that the new method is much more accurate than RANS. When the same mesh resolution is used, the new method is also more accurate than LES with the Smagorinsky (Smagorinsky 1963), k-equation-transport (Nicoud & Ducros 1999), or WALE (Kim et al. 1997) subgrid model.

The benchmark studies in Jin (2019) show the potential of the new method for simulating the engineering of turbulent flows with a complex geometry and a high Reynolds number. However, it is still not clear why the new method has a better performance than the other RANS and LES models. The purpose of this study is to better understand the mechanism of this method. The turbulent channel flows and decaying homogeneous and isotropic turbulence are simulated to demonstrate how this turbulence model works. Based on our study, we will try to find out how to develop more accurate and efficient turbulence models.

## 2. Mathematical model and numerical methods

For an incompressible flow, the governing equations of a generic turbulence model can be written as follows

$$\frac{\partial u_i}{\partial x_i} = 0, \tag{1}$$

$$\frac{\partial u_i}{\partial t} + \frac{\partial (u_i u_j)}{\partial x_j} = -\frac{\partial p}{\partial x_i} + \nu \frac{\partial^2 u_i}{\partial x_j^2} + g_i + \mathcal{M}_i. \tag{2}$$

$\mathcal{M}_i$ is a generic modeling term, which can be the modeled Reynolds stress term for RANS or the SGS term for LES.

In this study, we interpret turbulence modeling differently from RANS or LES. In this interpretation, the model solution is seen as an asymptotic approximation of the exact solution of turbulent flows (which can be seen as an ideal DNS solution with the mesh size $\Delta \rightarrow 0$ and time step $\delta t \rightarrow 0$). $\mathcal{M}_i$ is an artificial force which dissipates turbulent motions. Compared with DNS, a lower mesh resolution can be used in a model simulation if small eddies are dissipated. Thus, the purpose of turbulence modeling is to dissipate more small eddies without changing the main statistical field qualitatively, leading to the small-eddy-dissipation (SED) mechanism for turbulence modeling. We expect that more turbulent kinetic energy can be dissipated where the turbulence is stronger. Thus, the characteristic length scale of the modeling term $\mathcal{M}_i$ should be the length scale of turbulence $l_t$.

Classic LES models try to resolve large eddies directly and model small eddies, instead of dissipate small eddies. In order to model the SGS motions, most LES models use the mesh size $\Delta$ as the characteristic length scale, instead of $l_t$. We can see that, different turbulence models are derived because of the different interpretations of the physics.

2.1 Small-eddy-dissipation mixing length (SED-ML) model

Interpreting the ML+ model in Jin (2019) with the SED mechanism and making further improvements, we propose the following SED-ML model. $\mathcal{M}_i$ is still modeled as the product of a dissipation strength indicator $\phi$ and a dissipative force distribution $F_i$, i.e., $\mathcal{M}_i = \phi F_i$. $F_i$ is proposed based on the EVA, expressed as

$$F_i = \frac{\partial}{\partial x_j}\left(\nu_{eff}\frac{\partial u_i}{\partial x_j}\right). \tag{3}$$

It can be noticed that $\frac{\partial u_i}{\partial x_j}$, instead of the strain rate $2s_{ij} = \left(\frac{\partial u_i}{\partial x_j} + \frac{\partial u_j}{\partial x_i}\right)$, is used in equation (3). We don't need to use a symmetric tensor here, because $\mathcal{M}_i$ is interpreted as an artificial dissipation force, instead of the Reynolds stress tensor term or the residual stress tensor term. This will benefit the numerical simulation because calculation of the nonlinear term $\frac{\partial}{\partial x_j}\left(\nu_{eff}\frac{\partial u_j}{\partial x_i}\right)$ can be avoided.

According to the SED mechanism, $\mathcal{M}_i$ should be proportional to the strength of turbulence. We can use a modified mixing length $l'_{mix}$ to approximate the turbulence length scale and $\vartheta_{mix} = l'_{mix}|s_{ij}|$ as the characteristic velocity, where $|s_{ij}| = \left(2s_{ij}s_{ij}\right)^{1/2}$ is the magnitude of the transient strain rate $s_{ij}$. Then the effective viscosity $\nu_{eff}$ in equation (3) is calculated as

$$\nu_{eff} = l'^2_{mix}|s_{ij}|. \tag{4}$$

To propose $l'_{mix}$, we introduce the following local transient length scales,

$$l_l = \sqrt{\frac{|\partial K/\partial t|}{|s_{ij}|^3}}, \tag{5}$$

$$l_s = \sqrt{\frac{\nu}{|s_{ij}|}}, \tag{6}$$

where $K = \frac{1}{2}u_k^2$ is the instantaneous kinetic energy. $l_l$ characterizes the length scale of transient velocity fluctuation. $l_s$ is a small length scale, which is identical to the viscous length scale $\frac{\nu}{u_\tau}$ at the wall. $l_l$ ad $l_s$ can be calculated from the transient flow field directly, without using any statistical results. Using $l_l$, $l_s$, and the wall distance $y_{\mathrm{w}}$, we can establish the following local transient dimensionless numbers,

$$y_s^+ = \frac{l_l}{l_s} = \sqrt{\frac{|\partial K/\partial t|}{\nu |s_{ij}|^2}}, \tag{7}$$

$$y_\infty^+ = \frac{y_w}{l_s} = y_w \sqrt{\frac{|s_{ij}|}{\nu}} \tag{8}$$

While $y_{\mathrm{s}}^+$ characterizes the strength of transient fluctuation, $1/y_\infty^+$ characterizes the wall effect. Similar to the classic mixing length, $l'_{mix}$ is calculated as

$$l'_{mix} = \kappa y_w F_1(y_s^+) F_2(y_\infty^+). \tag{9}$$

$F_1(y_s^+)$ damps the mixing length near the wall, expressed as,

$$F_1(y_s^+) = 1 - exp(-y_s^+/A^+), \tag{10}$$

$A^+ = 1$ is a model constant. It can be seen that calculation of the friction velocity $u_\tau$ in the damping function proposed by Driest (1956) has been avoided in the damping function (10). $F_1(y_s^+)$ becomes zero not only at the wall but also when the flow is steady, so the model is also valid for steady laminar flows.

$F_2(y_s^+)$ damps the mixing length in the region far away from the wall, expressed as

$$F_2(y_\infty^+) = 1 - exp\left(-\left(B^+/y_\infty^+ + o(1/y_\infty^+)\right)\right). \tag{11}$$

There are the scaling laws $F_2(y_\infty^+) \sim 1$ as $y_w$ approaches zero, and $F_2(y_\infty^+) \sim B^+ l_s y_w^{-1}$ as $y_w$ approaches infinity. So, when the flow is far away from a wall, the mixing length is simplified as

$$l'_{mix} = \kappa B^+ l_s F_1(y_s^+), \tag{12}$$

which is independent of the wall distance $y_w$.

The high order terms $o(1/y_\infty^+)$ in equation (11) can be neglected as $y_w$ approaches infinity, however, it might affect the solution when the wall still has an effect. It makes $F_2(y_\infty^+)$ to approach one faster with a decrease of $y_\infty^+$. We are still unable to determine these high order terms. As an approximation, we assume $B^+/y_\infty^+ + o(1/y_\infty^+)$ is very large for wall bounded flows, leading to $F_2(y_\infty^+) = 1$. The model constant $B^+$ will be studied using the test case of decaying homogenous and isotropic turbulence, which is not affected by walls.

If we set the value of $\phi$ to 1, the proposed model becomes similar to a classic mixing length model. However, we only use small $\phi$ values in our simulation because the model solution of equations (1)-(2) is an asymptotic approximation of the exact solution, while $\phi$ acts as a small perturbation parameter. We set the default value $\phi_0$ to 0.004. It yields accurate solutions for decaying homogeneous and isotropic turbulence, smooth wall channel flows, rough-wall channel flows, and compressor cascade flows.

## 2.2 Parameter extension

The accuracy of the model solution can be further improved by using a linear extension. The corresponding simulation (a model solution with a parameter extension) is called PES in Jin (2019). This method originates from an asymptotic method suggested by Carey & Mollendorf (1980). Herwig and his colleagues further extended the method and performed comprehensive studies with respect to variable property effects on flow and heat transfer problems, see Herwig (1985; 1987); Bünger & Herwig (2009) and Jin & Herwig (2012) as examples. Hereby, we will analyze why a linear extension can improve the simulation accuracy of turbulent flows.

Besides the artificial dissipation $\phi F_i$, a numerical simulation with the time step $\delta t$ and mesh size $\Delta$ is also affected by a numerical dissipation. The numerical solution $R(\phi, \delta t, \Delta)$ of equations (1) and (2) approaches the exact solution $R_E(\phi = 0, \delta t = 0, \Delta = 0)$ as $\phi$, $\delta t$ and $\Delta$ approach zero. Thus, the numerical solution of equations (1)-(2) can be looked upon as an asymptotic approximation of the exact solution of the Navier-Stokes equations with $\phi$, $\delta t$ and $\Delta$ as small regular perturbation parameters.

To understand how a model solution $R$ approaches the exact solution $R_E$, we first think about an ideal simulation which has $\Delta = 0$ and $\delta t = 0$. When $\phi$ is small enough, the simulation error $\delta_{sim}(\phi) = R(\phi) - R_E$ approaches zero monotonically with the reduction of $\phi$, see curve A-B-C-D in figure 1. Taking only the leading order term, $\delta_{sim}$ can be approximated as

$$\delta_{sim}(\phi) \approx A\phi^n, \tag{13}$$

where $n$ is the leading scaling order. If we know two ideal model solutions for $\phi = \phi_0$ and $t_\phi \phi_0$, where $0 < t_\phi < 1$, we are able to approximate the exact solution by a linear extension, calculated as

$$R^e = R(\phi_0) - \frac{R(t_\phi \phi_0) - R(\phi_0)}{t_\phi - 1}. \tag{14}$$

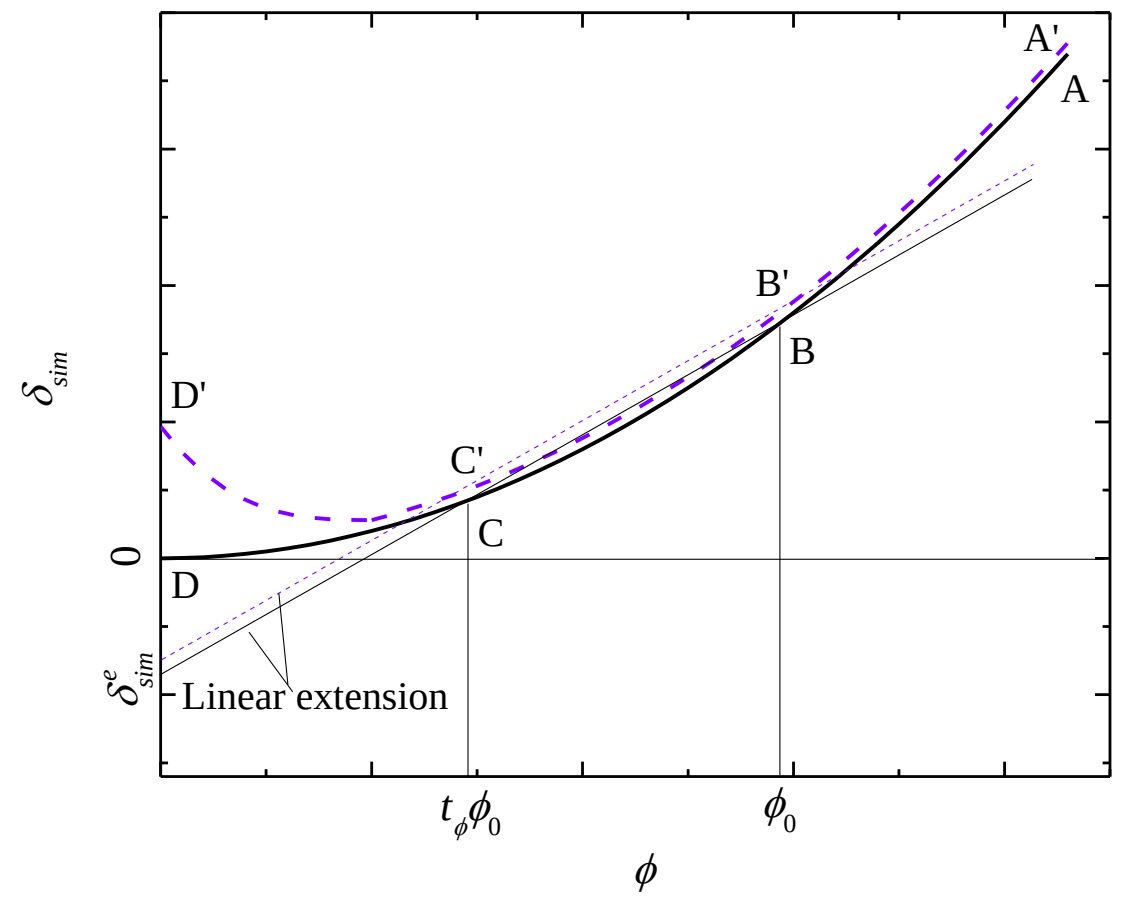

Figure 1 Schematic diagram for the relationship between $\phi$ and the simulation error $\delta_{sim}(\phi)$. Ideal solutions for $\delta t \rightarrow 0$ and $\Delta \rightarrow 0$ are indicated by the cure A-B-C-D. Real solutions for $\delta t > 0$ and $\Delta > 0$ are indicated by the cure A'-B'-C'-D'.

Subtracting $R_E$ from equation (14) and considering equation (13), the error of the extended solution $\delta^e_{sim} = R^e - R_E$ can be estimated as

$$\delta^e_{sim} = R^e - R_E = \delta_{sim}(\phi_0)\frac{t_\phi^n - t_\phi}{1 - t_\phi}. \tag{15}$$

Figure 2 shows the relationship between $\delta^e_{sim}/\delta_{sim}(\phi_0)$ and the scaling order $n$ for different values of $t_\phi$. It can be seen that the linear extension has the best performance when $n$ is close to 1. $\delta^e_{sim}$ has a sign opposite to $\delta_{sim}(\phi_0)$ when $n$ is larger than 1. When $\delta_{sim}(\phi)$ is only mildly non-linear ($n \leq 2$) and $t_\phi$ is set to be smaller than 0.8, the linear extension always improves the simulation accuracy, i.e., $|\delta^e_{sim}| < |\delta_{sim}(\phi_0)|$. $|\delta^e_{sim}/\delta_{sim}(\phi_0)|$ approaches its maximum value $\frac{t_\phi}{1-t_\phi}$ as $n \rightarrow \infty$. This means that the linear extrapolation might result into significant simulation error when the $\delta_{sim}(\phi)$ is strongly non-linear and $t_\phi$ is too large.

The curve A'-B'-C'-D' in figure 1 indicates schematically the relationship between $\delta_{sim}$ and $\phi$ for a real simulation with $\Delta > 0$ and $\delta t > 0$. We expect the numerical error (due to $\Delta$ and $\delta t$) and the explicit modeling term $\phi F$ will result in the simulation errors with the same signs, because both of them introduce artificial dissipations. So, we the curve A'-B'-C'-D' is above A-B-C-D, see figure 1.

A computational mesh might be insufficient for DNS ($\phi = 0$), leading to the mesh-dependence problem, see points D and D'. However, it is likely that the same mesh is sufficient for a model simulation with $\phi > 0$, because the modeling term $\phi F$ dissipates small eddies and they don't need to be resolved any more. Thus, we can make a similar linear extension for a real simulation,

$$R^e(\Delta, \delta t) = R(\phi_0, \Delta, \delta t) - \frac{R(t_\phi \phi_0, \Delta, \delta t) - R(\phi_0, \Delta, \delta t)}{t_\phi \phi_0 - \phi_0} \phi_0. \quad (16)$$

If the solutions $R(\phi_0, \Delta, \delta t)$ and $R\big(t_\phi \phi_0, \Delta, \delta t\big)$ are generally mesh independent, $R^e(\Delta, \delta t)$ is close to $R^e$, see the dotted line in figure 1. Therefore, we are still able to use a linear extension to correct the model solution.

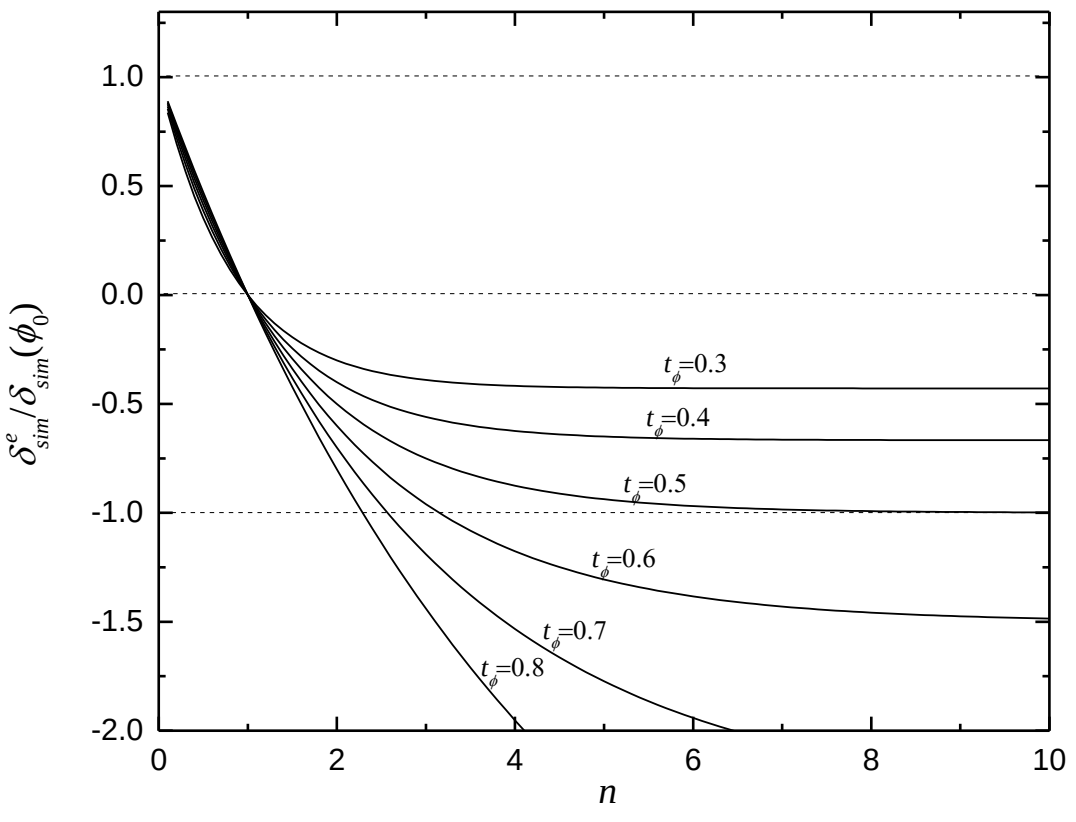

Figure 2 leading scaling order $n$ versus $\delta^e_{sim}/\delta_{sim}(\phi_0)$ for different values of $t_\phi$.

In a LES with near wall resolution, the mesh size $\Delta$ can be also seen as a dissipation strength indicator because the artificial dissipation also decreases with a decrease of $\Delta$. However, the linear extension cannot be applied to a LES. One reason is that it is hard to reduce $\Delta$ uniformly in a complex geometry. In addition, the solution $R$ as a function of $\Delta$ is often strongly non-linear.

## 2.3 Numerical method

For the simulations, a finite volume-method (FVM) was utilized. The solver was developed based on our DNS solver by using the open source code package OpenFoam. The spatial discretization was implemented by a second-order central-difference scheme. For time derivatives, the second-order implicit backward method was used. For the correction and coupling of the pressure and velocity fields the Pressure-Implicit scheme with Splitting of Operators (PISO) algorithm was used (Versteeg & Malalasekera 2007). A stabilized preconditioned (bi-)conjugate gradient solver was utilized to solve the pressure field and the momentum and species concentration equations. We have performed the code validation for our solver extensively in our previous studies (Jin *et al.* 2015; Uth *et al.* 2016; Jin & Kuznetsov 2017; Gasow *et al.* 2020; 2021; Jin 2019; 2020).

## 3. Application to smooth wall channel flows

To demonstrate how the SED-ML model with a parameter extension works, we have calculated the turbulent channel flows for $Re_\tau = 821$ and 2003. The Reynolds number $Re_\tau$ based on the friction velocity $u_\tau$ is defined as

$$Re_\tau = \frac{u_\tau H}{\nu}, \tag{17}$$

where $H$ is the half channel height. The size of the computational domain is $2\pi H \times 2H \times \pi H$.

The reference mesh (mesh A) has the resolution $96 \times 96 \times 96$. The mesh cells are uniformly distributed in the streamwise and spanwise directions, while they are concentrated near the wall in the wall-normal direction. With a constant factor $t_m = 1.5$, the number of mesh cells is increased uniformly three times in each direction, to understand how the numerical solution changes with the mesh resolution. The detailed information of the meshes in this study is shown in table 1.

With a constant factor $t_\phi = 2/3$, the dissipation strength indicator $\phi$ in the SED-ML model is decreased from $\phi_0 = 0.004$ to $t_\phi \phi_0 = 0.0027$ and $t_\phi^2 \phi_0 = 0.0018$, to understand how the model solution changes as $\phi$ approaches zero. Besides the SED-ML model, the test cases are also calculated using LES with and without subgrid models for comparison. The simulation results are validated with the DNS data in Kis (2011) and Hoyas & Jimenez (2008).

Table 1 Mesh resolutions used in the study. The number of mesh cells in each direction is increased uniformly three times with a factor $t_m = 1.5$.

| Mesh ID | Resolution | $\Delta y_w^+$ ($Re_\tau = 821$) | $\Delta y_w^+$ ($Re_\tau = 2003$) |
|---|---|---|---|
| A | $128 \times 128 \times 128$ | 2.67 | 4.00 |
| B | $192 \times 192 \times 192$ | 1.09 | 2.66 |
| C | $288 \times 288 \times 288$ | 0.73 | 1.77 |
| D | $432 \times 432 \times 432$ | 0.61 | 0.85 |

### 3.1 Friction coefficient

Multiplying equation (2) with $u_i$, taking the time averaging and volume averaging in the computational domain, we have

$$\delta_d + \delta_m = 1, \tag{18}$$

where $\delta_d = \frac{\langle \varepsilon \rangle}{g u_m}$ and $\delta_m = \frac{\langle \overline{\tilde{u}_i \mathcal{M}_i} \rangle}{g u_m}$ are the ratios of the directly resolved loss $\langle \varepsilon \rangle = \nu \langle \overline{\frac{\partial u_i}{\partial x_j} \frac{\partial u_i}{\partial x_j}} \rangle$ and modeled loss $\langle \overline{\tilde{u}_i \mathcal{M}_i} \rangle$ to the input power $g u_m$, respectively. The operator $\langle\ \rangle$ denotes the volume averaging in the computational domain; $\bar{}$ denotes the time averaging. In a real simulation, however, the kinetic energy is also dissipated implicitly due to the numerical error. The ratio of the numerical dissipation $\delta_n$ to $gu_m$ is estimated as $1 - \delta_d - \delta_m$. Jin, et al. (2015) suggests to use $\delta_n$ as an error measure. $\delta_m + \delta_n$ indicates the ratio of the artificial dissipation rate to the total dissipation rate. It accounts for both the modeled dissipation (indicated by $\delta_m$) and the numerical dissipation (indicated by $\delta_n$).

Figure 3 shows the relationship between the friction coefficient $f$ and the artificial dissipation ratio $\delta_m + \delta_n$ for $Re_\tau = 821$. The SED-ML model is compared with three LES models. The SED-ML model solution is generally mesh-independent when mesh C is used. Using a higher mesh resolution doesn't further improve the simulation accuracy.

It is evident in figure 3 that, when the same mesh resolution is used, the SED-ML model is much more accurate than the Smagorinsky model and the k-equation model, while it has a similar accuracy as the WALE model. To ensure the relative simulation error $|f/f_{DNS} - 1| \leq 5\%$, the SED-ML model can dissipate 20% of $gu_m$ artificially (80% of $gu_m$ needs to be directly calculated), while the WALE model can dissipate only 10% of $gu_m$. Our numerical results show that the turbulent motions are more efficiently dissipated by the SED-ML model.

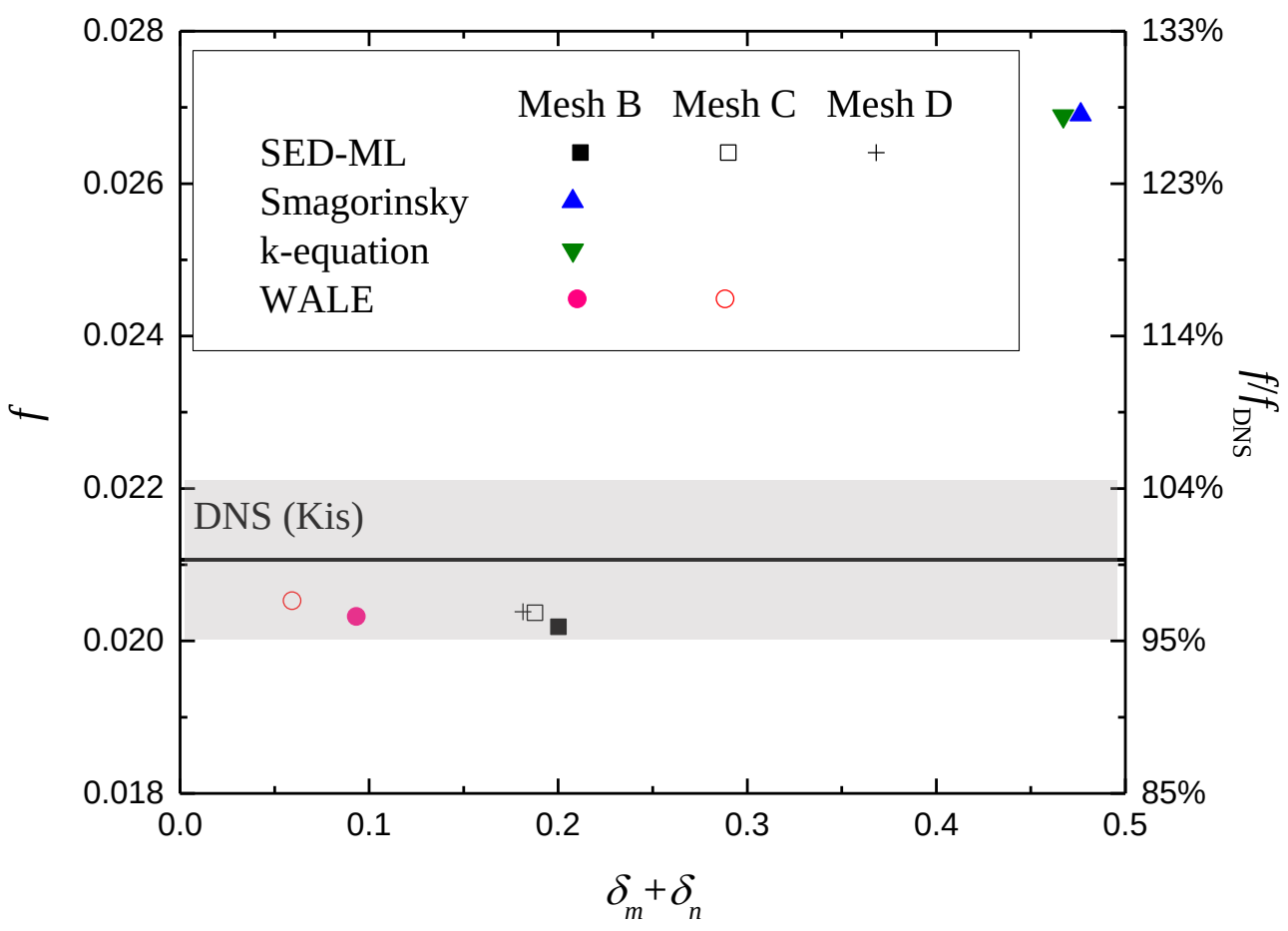

Figure 3. Ratio of artificial dissipation $\delta_m + \delta_n$ versus friction coefficient $f$ and $f/f_{DNS}$, $Re_\tau = 821$. The model results are compared with the DNS results of Kis (2011). The relative simulation error $|f/f_{DNS} - 1|$ is smaller 5% in the zone with grey color.

When the SED-ML model with mesh C is used in the case of $Re_\tau = 2003$, the relative simulation error is about 5%, see figure 4. The error decreases to 2% when mesh D is used. The solution of mesh C is slightly mesh-dependent. We don't pursue an absolutely mesh-independent solution. Instead, we only ensure that the influence of the numerical error on our numerical solution is negligibly small. This can be indicated by the magnitude of the error measure $\delta_n$.

Besides the WALE model, the LES without a subgrid model is also used in the comparison. Again, keeping the requested simulation error to 5%, the SED-ML model can dissipate 25% of $gu_m$ artificially, which is much higher than the other LES models.

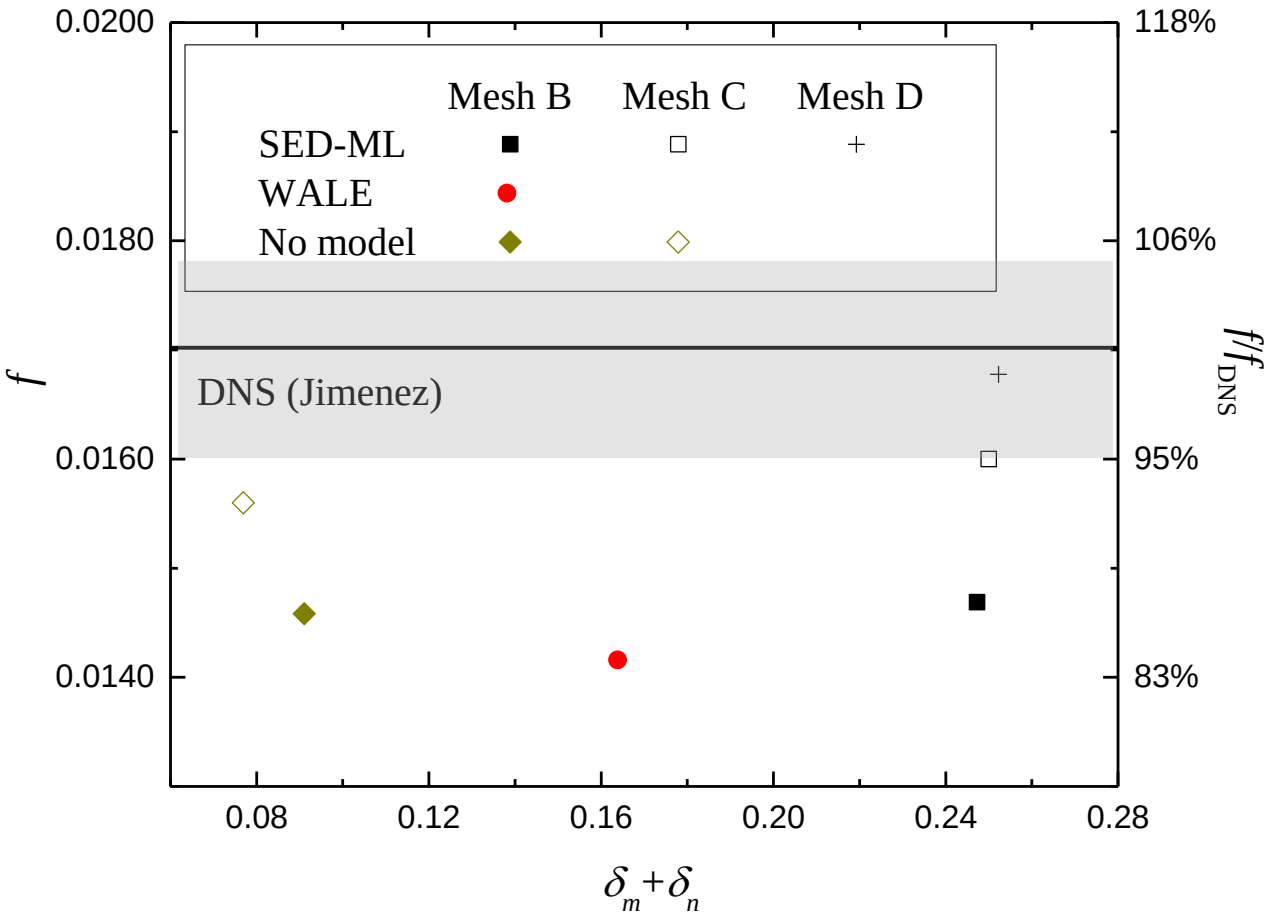

Figure 4. Ratio of artificial dissipation $\delta_m + \delta_n$ versus friction coefficient $f$ and $f/f_{DNS}$, $Re_\tau = 2003$. The model results are compared with the DNS results of Hoyas & Jimenez (2006). The relative simulation error $|f/f_{DNS} - 1|$ is smaller 5% in the zone with grey color.

We have carried out more simulations using the SED-ML model with different $\phi$ values. They are 0, $\phi_0$, $t_\phi \phi_0$, and $t_\phi^2 \phi_0$. The model solution of $f$ almost changes linearly with the reduction of $\phi$ when $\phi \geq t_\phi^2 \phi_0$, see figure 5. This linear relationship is broken when $\phi$ is close to 0 and the artificial dissipation is dominated by the numerical error. Using the solutions for $\phi_0$ and $t_\phi \phi_0$, we can make a linear extension to approximate the exact solution. Figure 5 shows that, a linear relationship with a similar slope holds when the modeled dissipation is dominating, even if the solution is still mesh-dependent.

Combing the solutions for high Reynolds numbers in this study and those for low and medium Reynolds numbers in Jin (2019), the friction coefficients predicted by the SED-ML model with and without a linear extension are compared with the DNS results in figure 6. While the SED-ML model is in reasonably accordance with the DNS results, the linear extension further improves the accuracy.

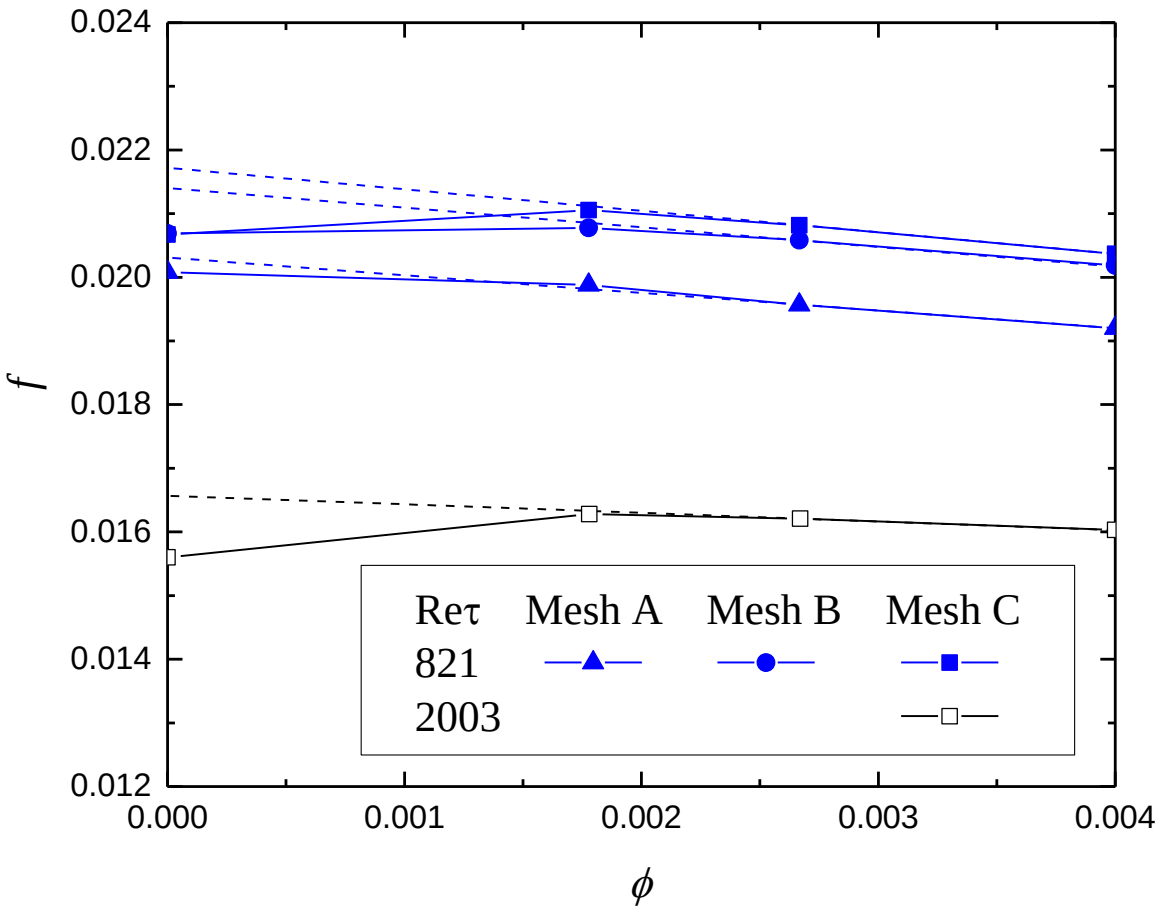

Figure 5. Dissipation strength indicator $\phi$ versus friction coefficient $f$. The linear extension, shown in dashed lines, is made using the solutions for $\phi_0$ and $t_\phi\phi_0$ with $t_\phi = 2/3$.

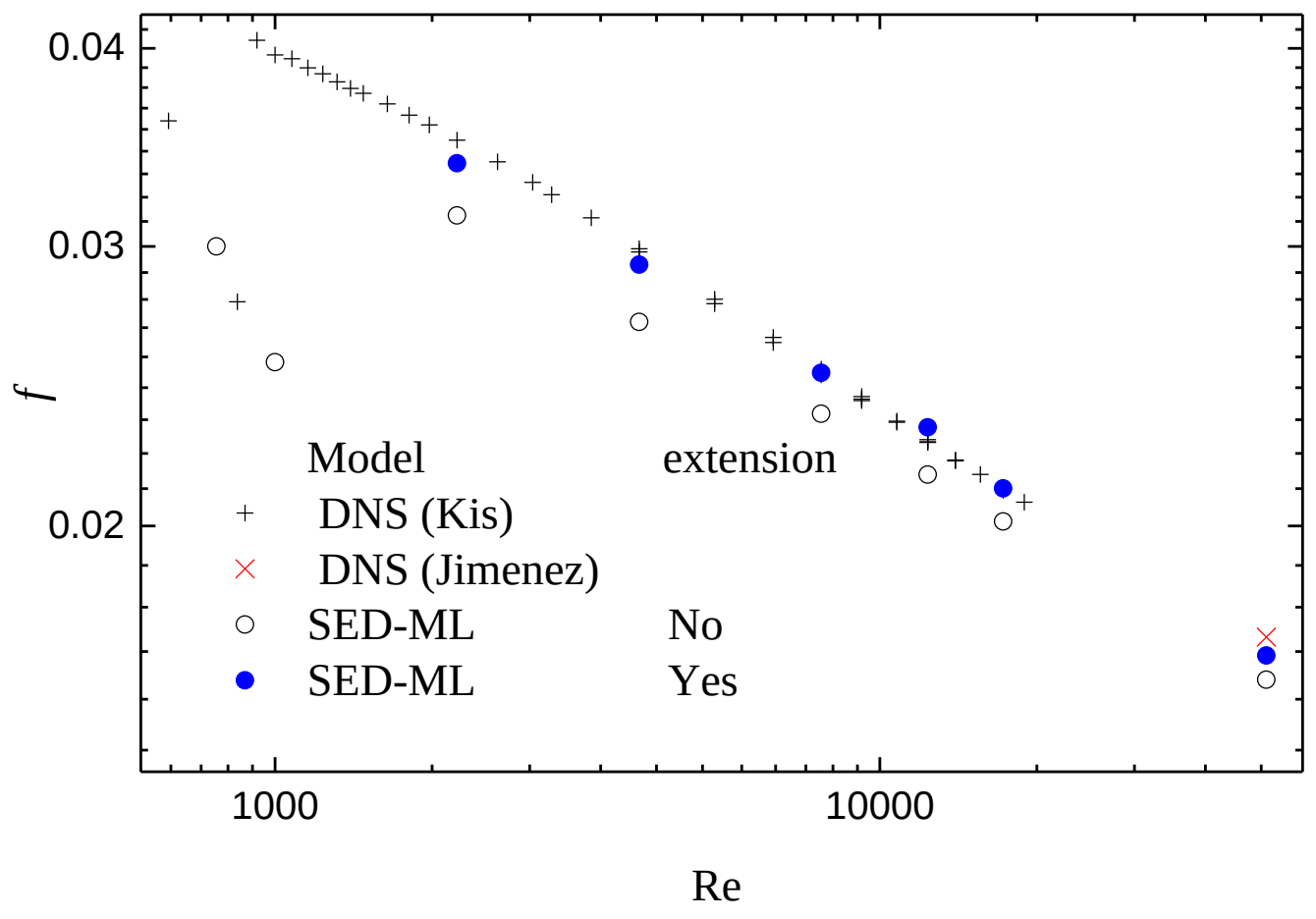

Figure 6. Friction coefficient $f$ versus Reynolds number $Re = \frac{u_m H}{\nu}$, where $u_m$ is the mean velocity. The model results are compared with the DNS results in Kis (2011) and Hoyas & Jimenez (2006).

### 3.2 Statistics of the velocity components

Figure 7 shows the distribution of the main velocity statistics in the wall-normal direction for $Re_\tau = 821$. They include the mean velocity $u_1^+$, r.m.s velocity components $u_1'^+$, $u_2'^+$ and $u_3'^+$, shear Reynolds stress $R_{12} = \overline{u_1' u_2'}$, and turbulent kinetic energy $k$. The superscript $^+$ denotes a quantity normalized with the friction velocity $u_\tau$ and viscous length scale $\frac{\nu}{u_\tau}$. The SED-ML solutions of all these quantities are in reasonable accordance with the DNS results, while the simulation accuracy can be further improved using the linear extension.

The model solutions of $u_1^+$, $u_2'^+$, $u_3'^+$, and $R_{12}$ for $Re_\tau = 2003$ are also improved by the linear extension, see figure 8. However, the linear extension doesn't evidently improve the solution of $u_1'^+$. One possible reason is that the numerical results for this high Reynolds number are still dependent of the mesh. Another reason could be that $u_1'^+$ has a strongly non-linear relationship with $\phi$ in the outer layer. Since the turbulent kinetic energy $k$ is a function of $u_1'^+$, $u_2'^+$ and $u_3'^+$, the calculated $k$ also deviates from the DNS results.

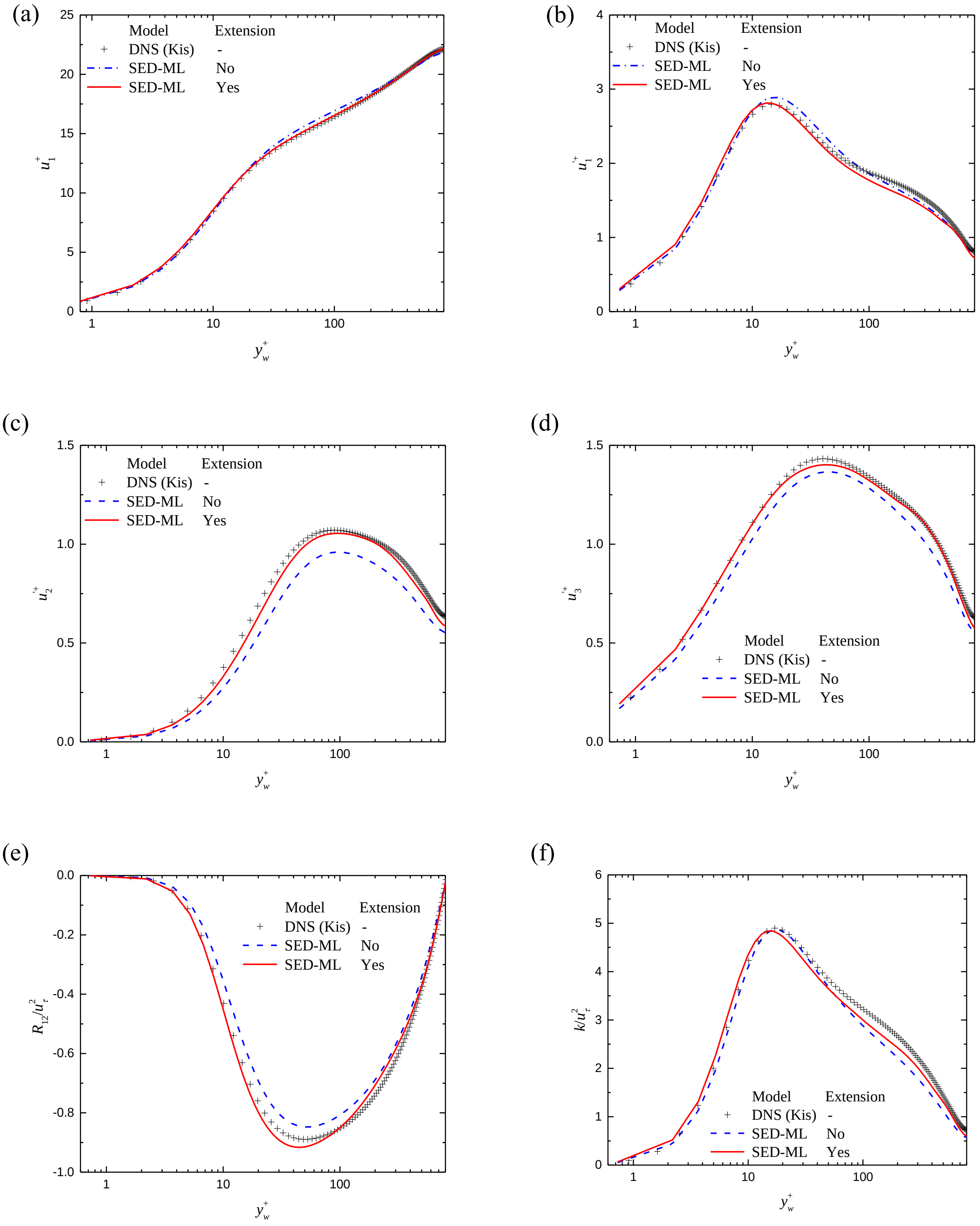

Figure 7. Distribution of the statistical results in the wall-normal direction for $Re_\tau = 821$, the model results are compared with the DNS data in Kis (2011), mesh C is used in the simulation. (a) $u_1^+$; (b) $u_1'^+$; (c) $u_2'^+$; (d) $u_3'^+$; (e) $R_{12}/u_\tau^2$; (f) $k/u_\tau^2$.

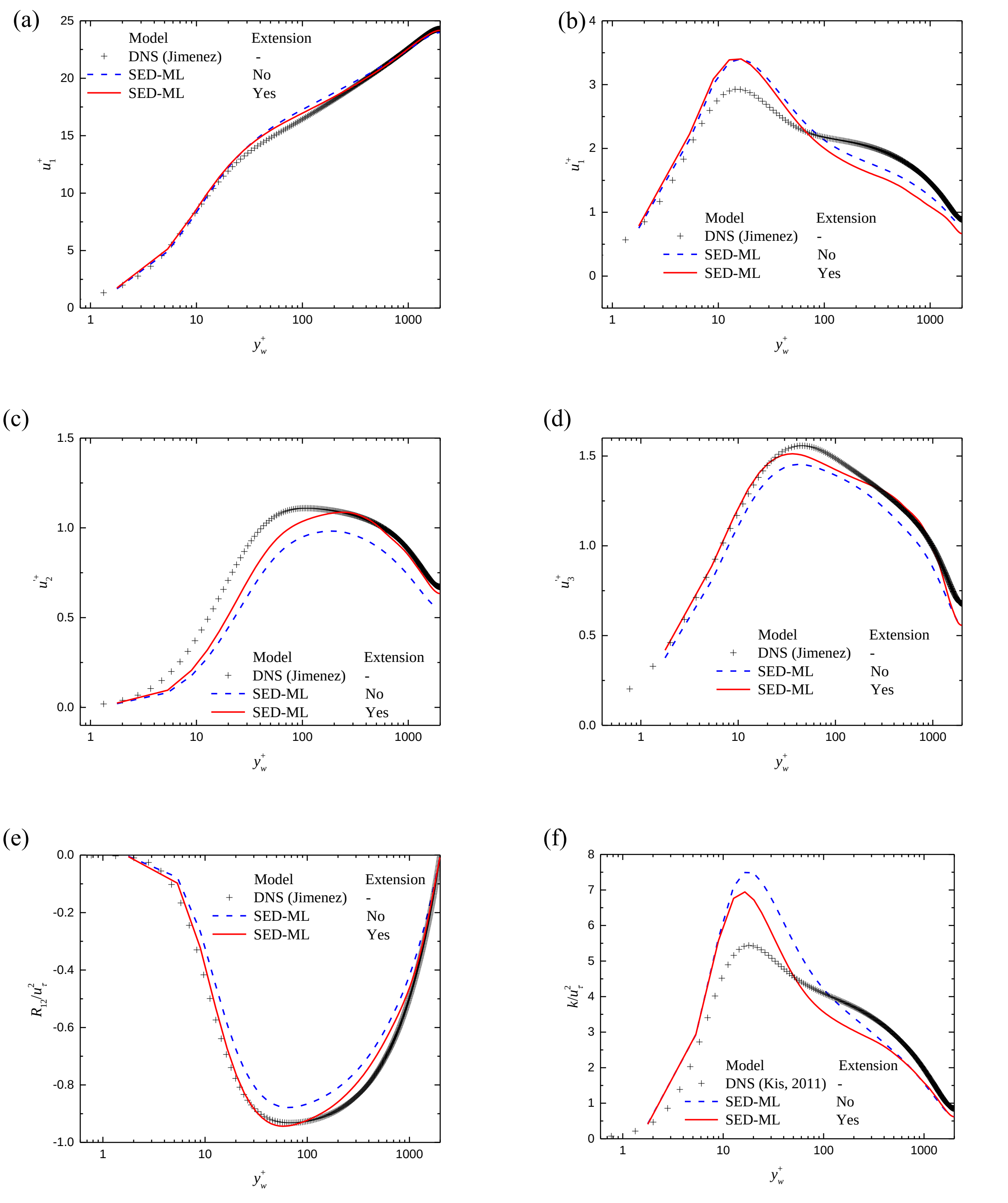

Figure 8. Distribution of the statistical results in the wall-normal direction for $Re_\tau = 2003$, the model results are compared with the DNS data in Kis (2011), mesh C is used in the simulation. (a) $u_1^+$; (b) $u_1'^+$; (c) $u_2'^+$; (d) $u_3'^+$; (e) $R_{12}/u_\tau^2$; (f) $k/u_\tau^2$.

### 3.3 Energy spectra

Figure 9 shows the energy spectra $\Phi_{11}$ and the pre-multiplied energy spectra $\kappa_1\Phi_{11}$ for the streamwise velocity fluctuation. In the viscous sublayer ($y^+ = 0.75$), $\Phi_{11}$ and $\kappa_1\Phi_{11}$ of the model solution are almost identical to those from DNS. This means that the SED-ML model dissipates very little kinetic energy when the turbulence is weak. It explains why the SED-ML model can capture the laminar-turbulent transition for compressor cascade flows, see Jin (2019; 2020).

By contrast, more kinetic energy is dissipated in the region away from the wall. The pre-multiplied energy spectra show that some large eddies are also dissipated by the artificial dissipation force. We might assume tentatively that, to reach a mesh-independent solution, the motions with $\Phi_{11}/(u_1'^2 H) \geq 10^{-4}$ need to be correctly calculated. Then, at the layer $y^+ = 328$, the turbulent motions with $\kappa_1 H \leq 50$ should be resolved in DNS, while only turbulent motions with $\kappa_1 H \leq 35$ need to be resolve in the SED-ML model. The SED-ML model requires to resolve only larger eddies. This may explain why the SED-ML model requires a lower mesh resolution than DNS.

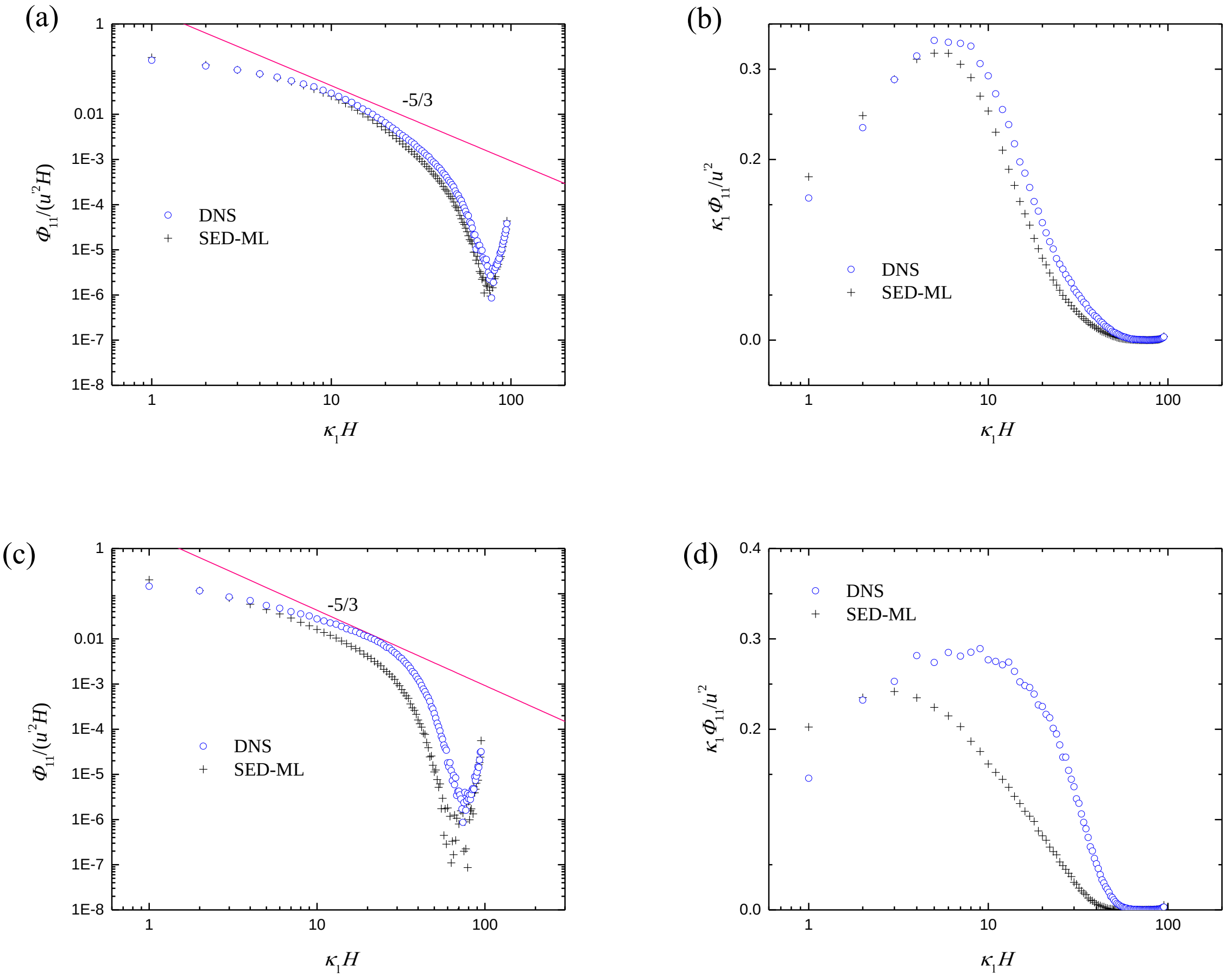

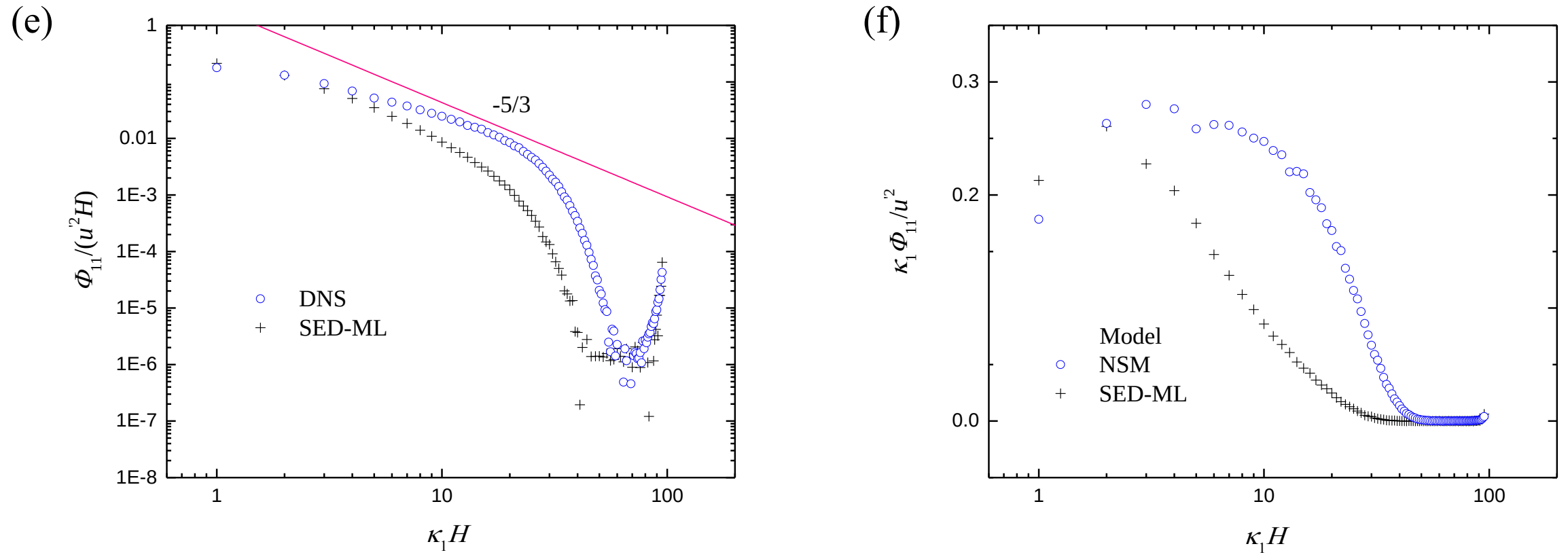

Figure 9. Energy spectra and premultiplied energy spectra at $y^+ = 0.75$ (a, b), 166 (c, d), and 328 (c, d) for $Re_\tau = 821$.

### 3.4 Budget of turbulent kinetic energy

The transport equation for the filtered turbulent kinetic energy can be derived from equations (1)-(2) as follows,

$$\frac{\partial k}{\partial t} + \frac{\partial(u_i k)}{\partial x_j} = \frac{\partial}{\partial x_i}\left(D_t + D_p + D_v\right) + Pro - \hat{\varepsilon} - \varepsilon_M, \tag{19}$$

where the turbulent, pressure, and viscous diffusion terms are

$$D_t = -\frac{1}{2}\frac{\partial \overline{u_i' u_j' u_j'}}{\partial x_i}, \tag{20}$$

$$D_p = -\frac{\partial \overline{p' u_i'}}{\partial x_i}, \tag{21}$$

$$D_v = \nu \frac{\partial^2 k}{\partial x_i^2}, \tag{22}$$

The turbulence production rate is

$$Pro = -\overline{u_\iota u_J}\frac{\partial \bar{u}_i}{\partial x_j}. \tag{23}$$

The pseudo turbulence dissipation rate is

$$\hat{\varepsilon} = \nu \overline{\frac{\partial u_i'}{\partial x_J}\frac{\partial u_i'}{\partial x_J}}. \tag{24}$$

$\hat{\varepsilon}$ and the turbulence dissipation rate $\varepsilon_D$ has the following relationship

$$\varepsilon = \hat{\varepsilon} + \nu \frac{\partial^2 \overline{u_i' u_J'}}{\partial x_i \partial x_j}. \tag{25}$$

The loss of turbulent kinetic energy due to $\mathcal{M}_i$ is

$$\varepsilon_M = \bar{\mathcal{M}}_i \bar{u}_i - \overline{\mathcal{M}_\iota u_\iota}. \tag{26}$$

The budget terms from the SED-ML model solution are compared with the DNS results, see figure 10. It can be seen that $Pro$ and $\varepsilon$ are under-predicted by the SED-ML model because some turbulent motions are dissipated artificially. However, the model results with the linear extension are in good accordance with the DNS results. Besides $Pro$ and $\varepsilon$, the diffusion terms $D_t$, $D_p$ and $D_v$ are also calculated with a reasonable accuracy.

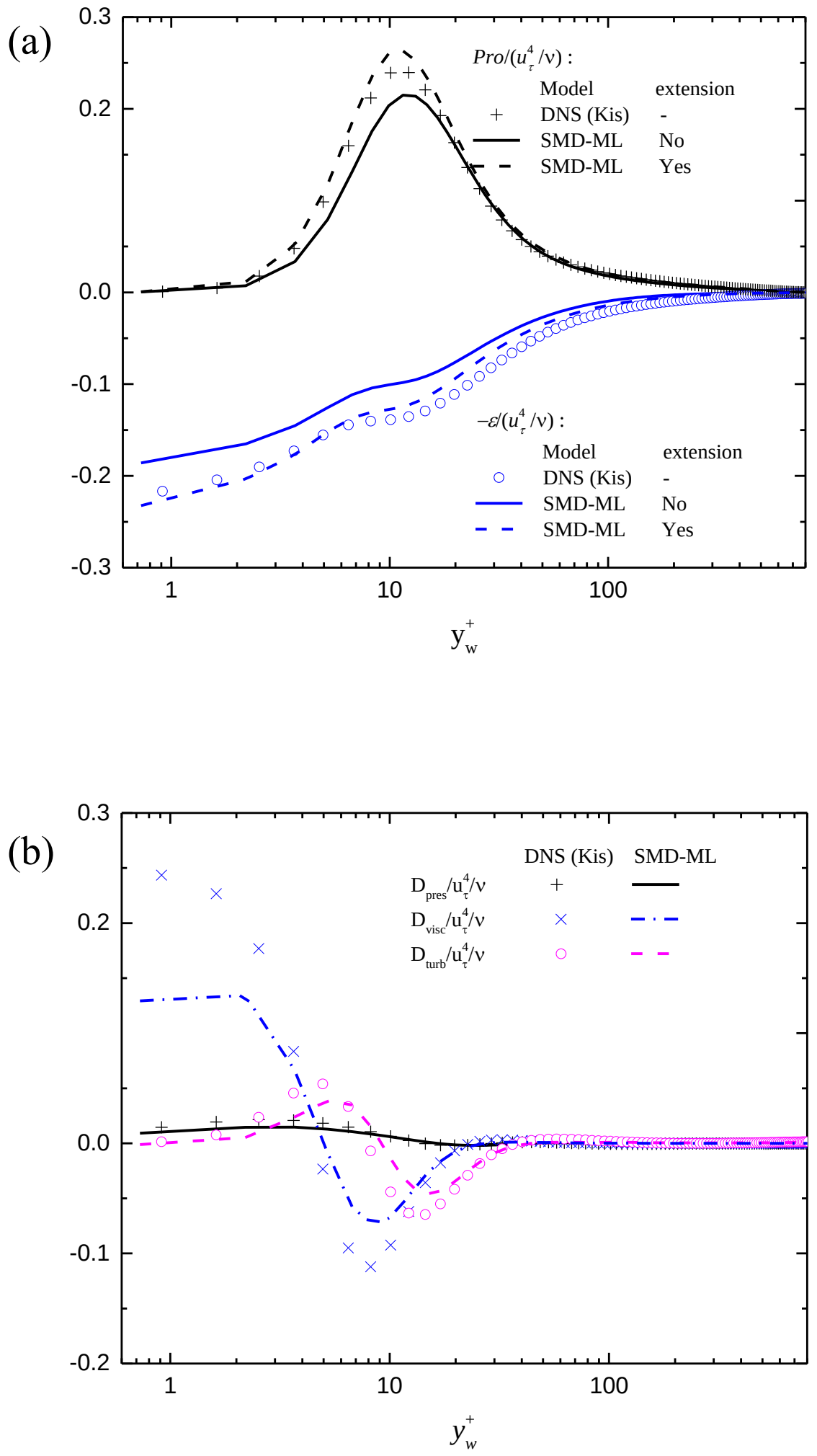

Figure 10. Budget of the turbulent kinetic energy for $Re_\tau = 821$. (a) Turbulence production rate $Pro$ and dissipation rate $\varepsilon$; (b) Pressure diffusion $D_{pre}$, viscous diffusion $D_{vis}$ and turbulent diffusion $D_{tur}$. The model results are compared with the DNS results by Kis (2011).

### 3.5 Loss of kinetic energy

When we propose the turbulence model, we try to dissipate the turbulent motions according to the strength of the turbulence. For this purpose, we use a modified mixing length $l'_{mix}$ to approximate the characteristic length of turbulence $l_t$. Normalized with the viscous length scale, the mean $l'_{mix}$ are compared with the distance from the wall $y_w$ and the characteristic length scale for a $k-\varepsilon$ model,

$$l_{k\varepsilon} = 0.09\frac{k^{3/2}}{\varepsilon}, \tag{27}$$

where the DNS results are used to calculate $k$ and $\varepsilon$. It can be seen that $l'^{+}_{mix}$ deviates from $y^+_w$ only in the viscous sublayer due to the damping function $F_1(y^+_s)$. A deficiency for using $l'_{mix}$ as the characteristic turbulence length scale is that $l'_{mix}$ approaches infinity with an increase of the wall distance $y_w$. If we set the model constant $B^+$ in the second damping function $F_2(y^+_\infty)$ to 20 and neglect the higher order terms $o(1/y^+_\infty)$, $l'^{+}_{mix}$ becomes close to $l^+_{k\varepsilon}$. However, we are still not able to determine the high order terms in $F_2(y^+_\infty)$ accurately. For simplicity, we still set $y^+_\infty$ to $+\infty$ for wall bounded flows.

The effects of $\mathcal{M}_i$ on the kinetic energy $E_M = -\overline{\mathcal{M}_i u_i}$ and on the turbulent kinetic energy $\varepsilon_M = \bar{\mathcal{M}}_i \bar{u}_i - \overline{\mathcal{M}_i u_i}$ are compared with the turbulent kinetic energy $k$ in figure 12. The similar trends of these quantities validates our assumption that the turbulent motions are generally dissipated according to the strength of the turbulence.

Due to the deficiency of $l'_{mix}$, $E_M$ is larger than $\varepsilon_M$ in the region $\frac{y_w}{H} > 0.6$, see figure 12, suggesting that that $\mathcal{M}_i$ not only dissipates turbulent kinetic energy $k$ but also the mean-flow kinetic energy $\bar{k} = \frac{1}{2}\bar{u}_i\bar{u}_i$. The model accuracy might be further improved by accounting for the higher order terms in $F_2(y^+_\infty)$. This will be our work in the future.

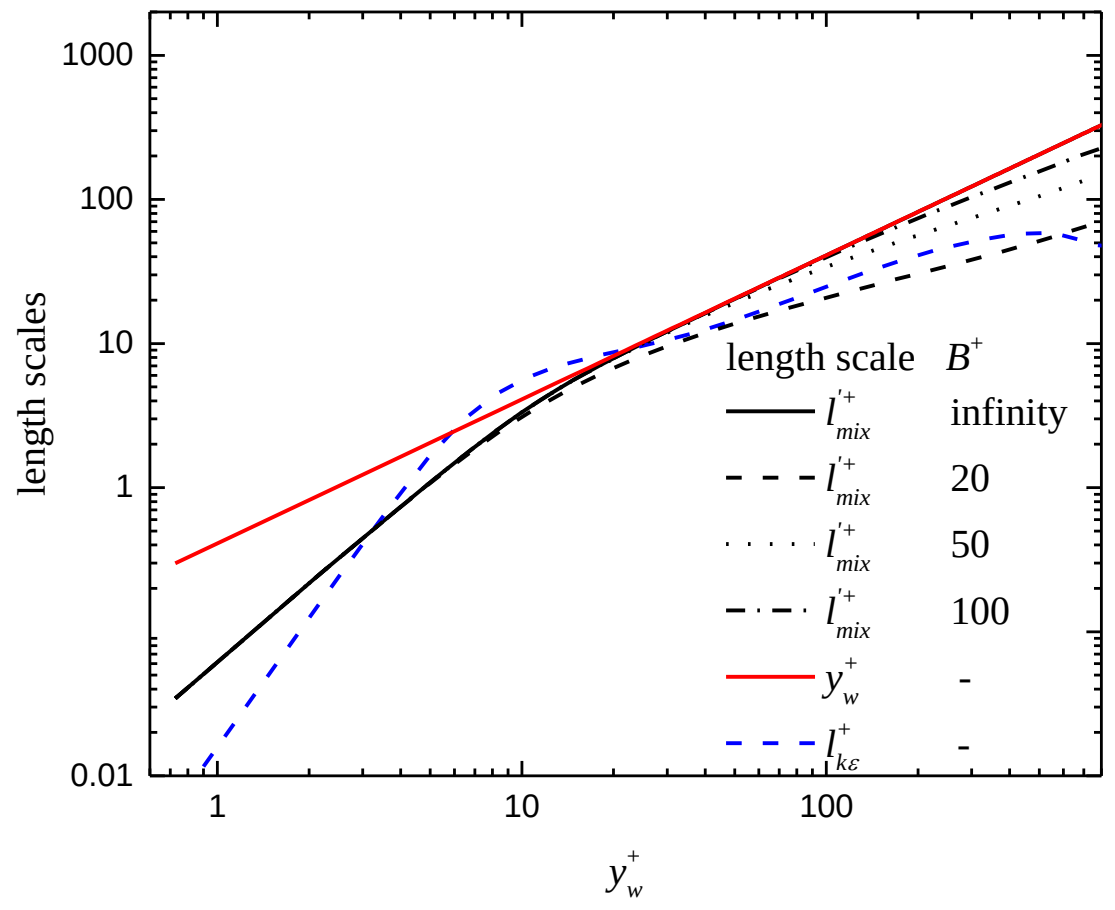

Figure 11. Length scales of turbulence, $Re_\tau = 821$.

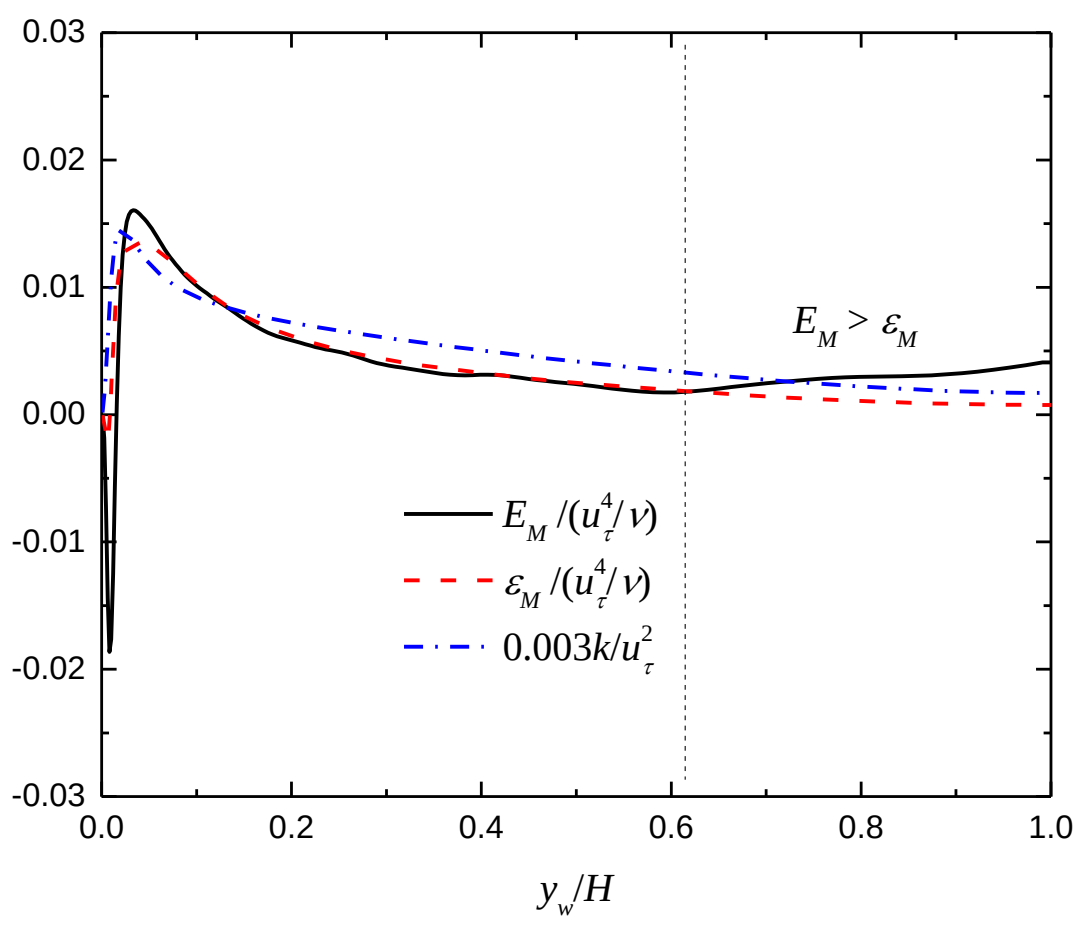

Figure 12. Losses of the mean kinetic energy $E_M = -\overline{\mathcal{M}_\iota u_\iota}$ and turbulent kinetic energy $\varepsilon_M = \bar{\mathcal{M}}_i \bar{u}_i - \overline{\mathcal{M}_\iota u_\iota}$ due to the turbulence modeling, compared with the turbulent kinetic energy $k$, $Re_\tau = 821$.

## 4. Application to decaying homogeneous and isotropic turbulence

To better understand the second damping function $F_2(y_\infty^+)$, we have calculated the homogeneous and isotropic turbulence in a box of size $2\pi \times 2\pi \times 2\pi$. This flow is not affected by walls, so damping function $F_2(y_\infty^+)$ in equation (9) is simplified to equation (12), which is for $y_w \to \infty$.

Periodic boundary conditions are given in all three directions. The Reynolds number based on the non-dimensional viscosity $\mathrm{Re} = 1/\nu$ is $10^5$. In the initial field, the velocity amplitude $Ek$ for the wave number $\mathbf{k} = (k_1, k_2, k_3)$ is given as

$$Ek(\mathbf{k}) = Ea \frac{|\mathbf{k}|^4}{k_0} \exp\left[-2\left(\frac{|\mathbf{k}|}{k_0}\right)^2\right], \tag{28}$$

where $Ea = 10$ and $k_0 = 5$ are constants. The mesh resolution, $128 \times 128 \times 128$, is used in the simulation.

Different from Jin (2019), this flow is calculated with the improved SED-ML model, with the default dissipation strength indicator $\phi_0 = 0.004$. The model constant $B^+$ is set to the value 20. With another solution for $\phi = \frac{2}{3}\phi_0 = 0.0027$, the model solution is corrected using a linear extension. Figures 13 and 14 show that, the kinetic energy is decaying fast because our initial field is different from real homogeneous and isotropic turbulence. While the SED-ML solution is in reasonable according with the DNS results, the linear extension further improves the simulation accuracy.

Because the mean velocity for this flow is zero, the volume averaged small length scale

$\langle l_s \rangle = \langle \sqrt{\frac{\nu}{|s_{ij}|}} \rangle$ is close to the Kolmogorov scale $\eta = \sqrt{\frac{\nu}{\langle s_{ij} \rangle}}$. When the flow is fully developed, there is $-dk/dt = \varepsilon$. So, the volume averaged dimensionless number $\langle y_s^+ \rangle = \langle \sqrt{\frac{|\partial K/\partial t|}{\nu |s_{ij}|^2}} \rangle$ is close to 1. Then, we can estimate the volume averaged mixing length

$$\langle l'_{mix} \rangle \approx \kappa B^+ \left(1 - \frac{1}{e}\right) \eta, \tag{29}$$

which has a linear relationship with the Kolmogorov scale $\eta$.

Figure 15 shows that $\langle l'_{mix} \rangle$ with $B^+ = 20$ is close to $l_{k\varepsilon}$. This is in accordance with our results for channel flows, i.e., $l'^+_{mix}$ with $B^+ = 20$ is close to $l^+_{k\varepsilon}$, see figure 11 for comparison. However, $\langle l'_{mix} \rangle$ increases with a decrease of $k$. The trend of $\langle l'_{mix} \rangle$ is different from $l_{k\varepsilon}$, but similar to $\eta$. The numerical results confirm the linear scaling $\langle l'_{mix} \rangle \sim \eta$ suggested by equation (29).

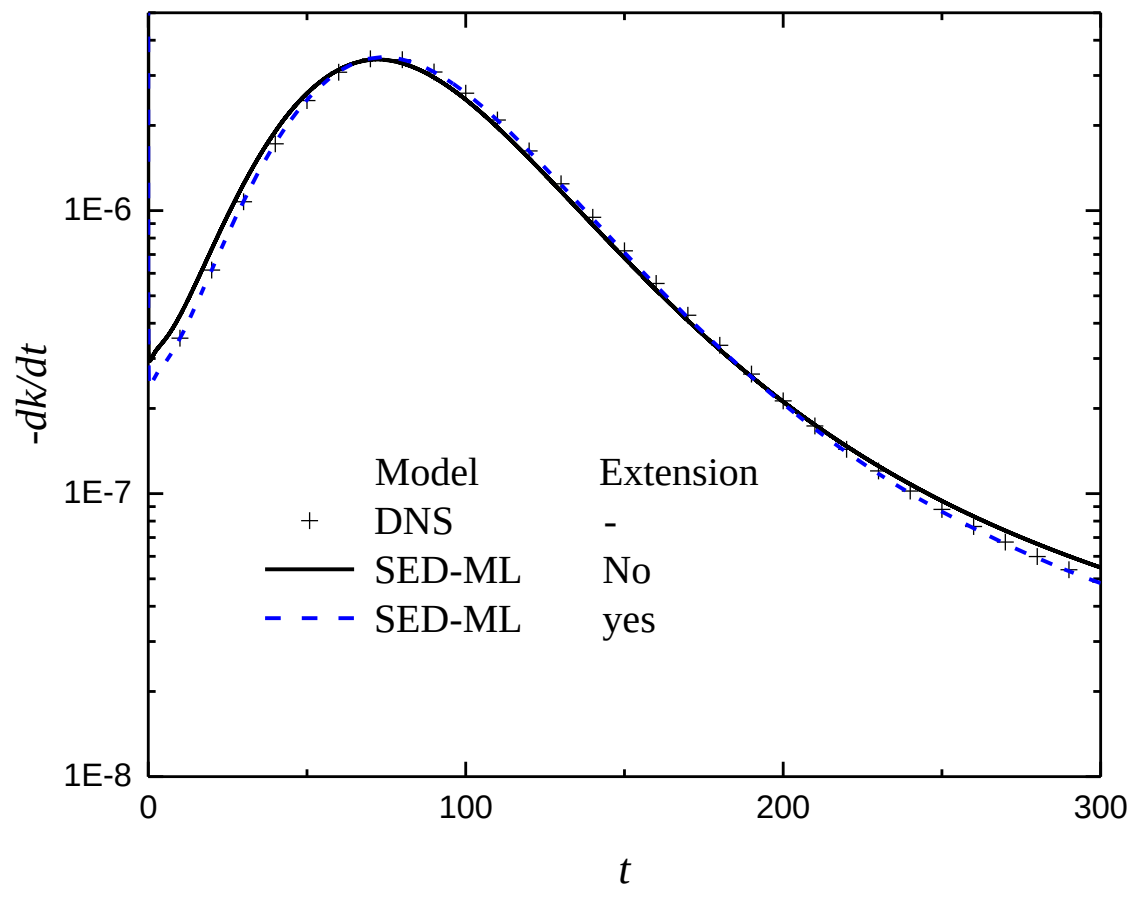

Figure 13. History of kinetic energy decaying rate $-dk/dt$, $B^+ = 20$.

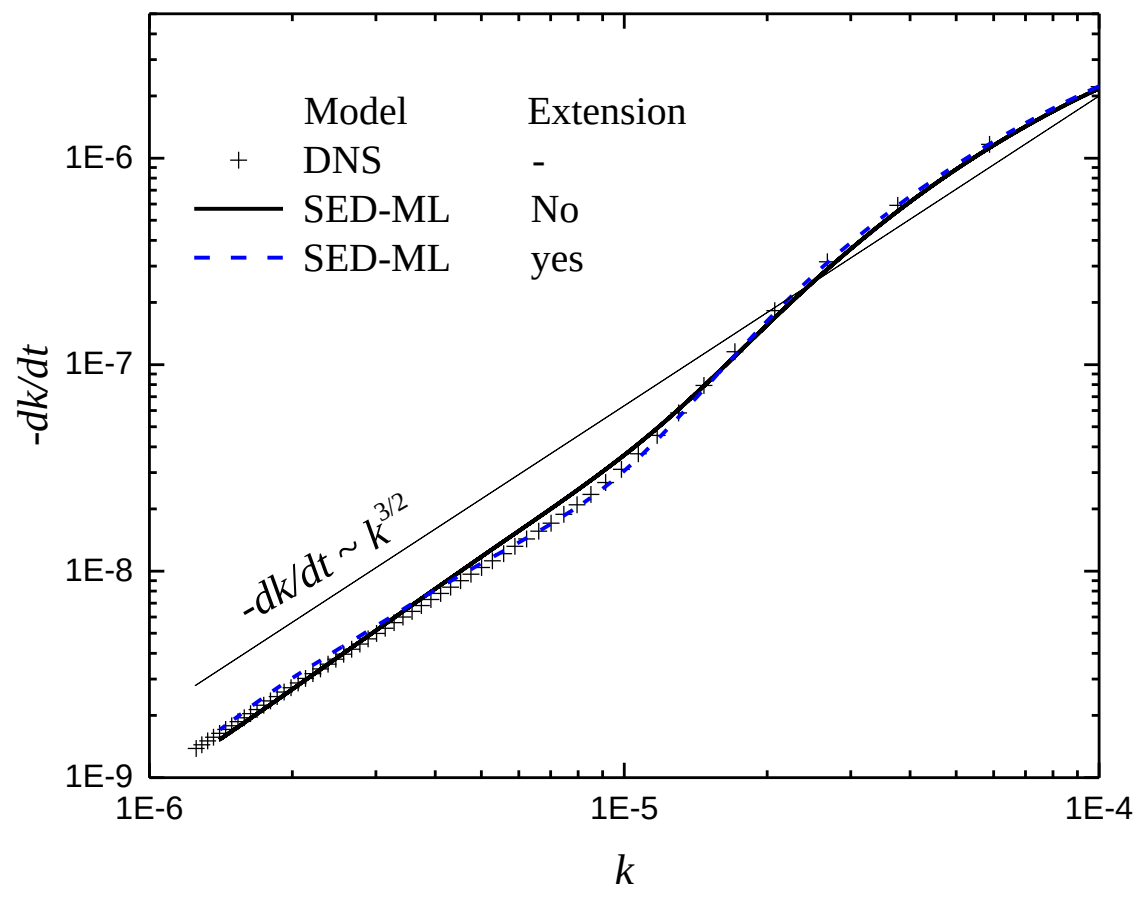

Figure 14. Kinetic energy decaying rate $-dk/dt$ versus kinetic energy $k$, $B^+ = 20$.

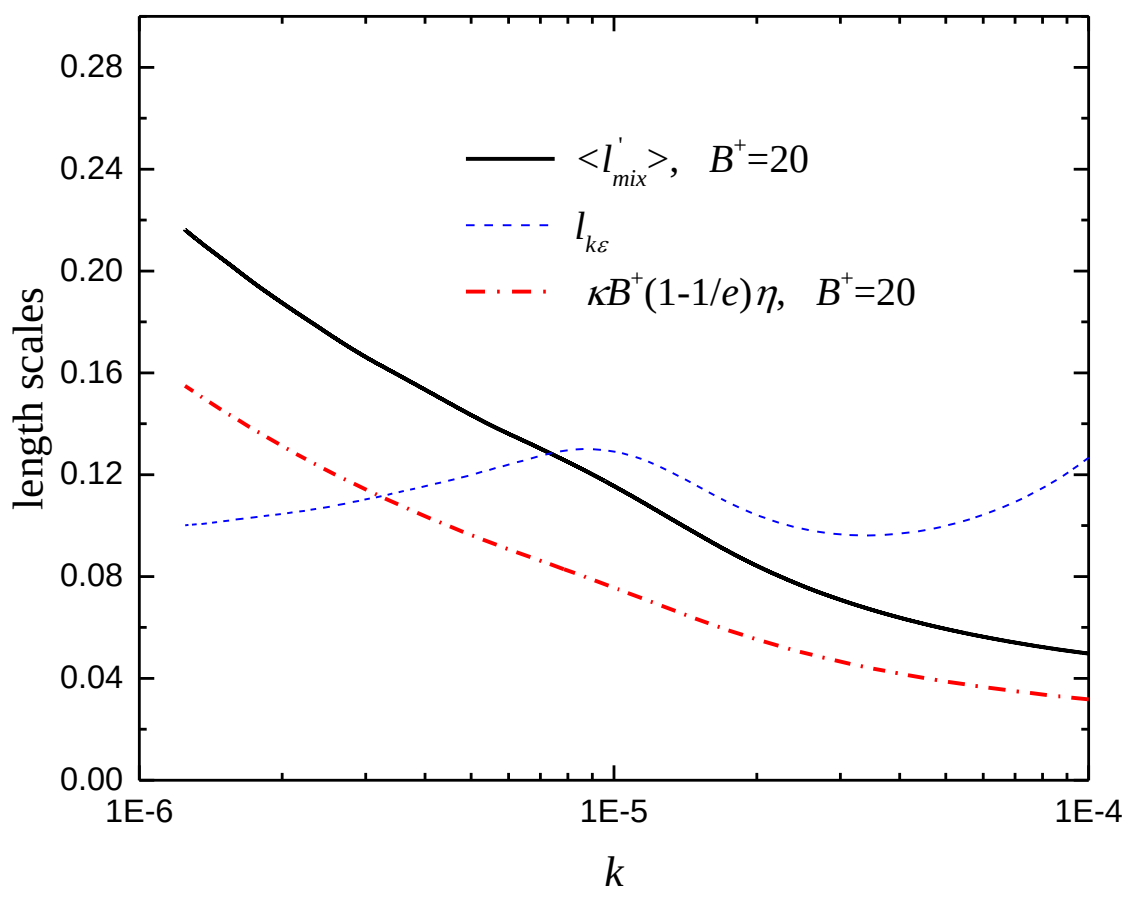

Figure 15. Length scales of turbulence $l'_{mix}$ and $l_{k\varepsilon}$ versus kinetic energy $k$, $B^+ = 20$.

**5. Discussions of numerical results**

Our numerical results in sections 3 and 4 confirm that the SED-ML model is in accordance with the SED mechanism: The model solution is qualitatively similar to the exact solution (identified by the DNS solution in this study). With the reduction of the artificial dissipation, the model solution approaches the exact solution asymptotically. The SED-ML model dissipates the kinetic energy according to the strength of the turbulence (turbulent kinetic energy), which means that the flow in the viscous sublayer is only slightly dissipated. This explains why the SED-ML model can capture the weak turbulence such as laminar-turbulent transition.

Because the kinetic energy is more efficiently dissipated, the SED-ML model is more accurate than a LES with the Smagorinsky model, k-equation model, or WALE model, when the same mesh resolution is used. Keeping the requested relative simulation error to be $\leq 5\%$, the SED-ML model can dissipate more kinetic energy than other LES models. That's why the SED-ML model requires a lower mesh resolution. The analysis also shows that the Smagorinsky model and k-equation model introduce too strong artificial dissipation, for this reason the friction coefficient is over-predicted by these models.

Most statistical results, including the friction coefficient, mean velocity, Reynolds stresses, turbulence production and dissipate rates, change monotonically with the reduction of the dissipation strength indicator $\phi$ when $\delta_m$ is dominating. The accuracy of these statistical results can be further improved by using a linear extension. The linear extension improve the simulation accuracy evidently at a low or medium Reynolds number, because the main statistical results change almost linearly with $\phi$. The accuracy is only mildly improved at a high

Reynolds number. A possible reason is that the length scale of turbulence $l_t$ is over-predicted in the outer layer by the current model. This deficiency might be solved by accounting for $O(1/y_\infty^+)$ in the second damping function $F_2(y_\infty^+)$.

## 6. Conclusions

We have proposed the SED mechanism for developing turbulence models in this study. According to this mechanism, the model solution is an asymptotic approximation of the exact solution of the Navier-Stokes equations. The turbulence modeling introduces an artificial dissipation (explicitly or implicitly) to dissipate small eddies. The simulation can be carried out using a lower-resolution mesh because the small eddies are dissipated. The purpose of turbulence modeling is to dissipate more small eddies without changing the statistical field qualitatively. For this purpose, the strength of artificial dissipation should be proportional to the strength of turbulence. So, the characteristic length scale used in a turbulence model should be the length scale of turbulence, instead of the mesh size as in a LES. The simulation accuracy can be further improved using a linear extension, when the solution $R$ changes monotonically with the reduction of a dissipation strength indicator ($\phi$ in this study), and the scaling law $R(\phi)$ is not strongly nonlinear.

The SED mechanism is different from RANS which approximates the Reynolds stresses and LES which approximates the sub-grid motions, while it is closer to the physics of turbulence modeling. Due to this reason, the new turbulence simulation method is more accurate than RANS and requires a lower computational cost than LES. According to the SED mechanism, a RANS solution is likely out of the monotonic interval, because it introduces too strong artificial dissipation (all turbulent kinetic energy has been dissipated). That is why it is hard (if possible) to propose a RANS model with high generality. Similarly, a LES model with a near wall treatment also introduces too strong dissipation near the wall. This might lead to uncertainties of the solution because the turbulence in the viscous sub-layer is rather weak.

The SED-ML model uses a modified mixing length $l'_{mix}$ to approximate the length scale of turbulence $l_t$. We expect there are two damping functions for $l'_{mix}$. They damp $l'_{mix}$ near the wall and in the region far away from the wall. We take the zero order term of the second damping function when wall bounded flows are simulated for simplicity. However, our numerical results near the channel center and for decaying homogeneous and isotropic turbulence show the necessity of using higher order terms of this damping function. We expect the simulation accuracy for flows with high Reynolds numbers can be further improved when $l_t$ is more precisely modeled. Despite of this deficiency, our analysis shows the potential of using the SED mechanism to develop turbulence models which have a high accuracy and a low computational cost.

**Acknowledgments**

The authors gratefully acknowledge the support of this study by the DFG-Heisenberg program (299562371), computing center of Hamburg University of Technology (RZ-TUHH), and the North-German Supercomputing Alliance (HLRN). The acknowledgement is also given to Prof. H. Herwig of Hamburg University of Technology for the insightful discussion with him.

**Declaration of Interests**

The authors report no conflict of interest.